\documentclass[conference]{IEEEtran}

\newcommand{\eg}{{\it e.g., }}
\newcommand{\etal}{{\it et~al.}}
\newcommand{\ie}{{\it i.e., }}

\usepackage{amsmath, scalerel}

\usepackage[linesnumbered,ruled]{algorithm2e}
\usepackage{graphicx}
\graphicspath{ {images/} }
\usepackage{pgfplots, lipsum}
\usepackage{textcomp}
\pgfplotsset{compat = newest}

\pgfplotsset{width=10cm,compat=1.9}
\usepgfplotslibrary{external}
\usepackage{tikz}
\usetikzlibrary{arrows}
\usepackage{caption}
\usepackage{siunitx}
\usepackage[font={small}]{caption}
\usepackage{amsmath}
\usepackage[export]{adjustbox}
\usepackage{lipsum}
\usepackage{subcaption}

\usepackage{xcolor}
\usepackage{pgfplots}
\usepackage{tikz}
\usepackage{pgfkeys}
\usepackage{url}
\usepackage{pgfplotstable}
\usepackage{subcaption}
\usepackage{filecontents}
\usepackage{multirow}

\makeatletter
\def\BState{\State\hskip-\ALG@thistlm}

\makeatother

\usepackage{cite}

\ifCLASSINFOpdf
  
\else
  
\fi

\begin{document}

\title{Cost Efficient Repository Management for Cloud-Based On-Demand Video Streaming 
}

\author{\IEEEauthorblockN{Mahmoud Darwich$^{*\ddag}$, Ege Beyazit$^{*}$, Mohsen Amini Salehi$^{\dag\ddag}$, Magdy Bayoumi$^{*}$}
\IEEEauthorblockA{ $^{*}$Center  for Advanced Computer Studies (CACS)\\
$^\ddag$High Performance Cloud Computing (HPCC) Laboratory\\
$^\dag$School of Computing and Informatics\\
University of Louisiana at Lafayette, Louisiana 70503\\
Email: \{mkd1007, exb6143, amini, mab\}@louisiana.edu}
}

\maketitle

\begin{abstract}
Video transcoding is the process of converting a video to the format supported by the viewer's device. Video transcoding requires a huge storage and computational resources, thus, many video stream providers choose to carry it out on the cloud. Video streaming providers generally need to prepare several formats of the same video (termed pre-transcoding) and stream the appropriate format to the viewer. However, pre-transcoding requires enormous storage space and imposes a significant cost to the stream provider. More importantly, pre-transcoding proven to be inefficient due to long-tail access pattern to video streams in a repository. To reduce the incurred cost, in this research, we propose a method to partially pre-transcode video streams and re-transcode the rest of it in an on-demand manner. We will develop a method to strike a trade-off between pre-transcoding and on-demand transcoding of video streams to reduce the overall cost. Experimental results show the efficiency of our approach, particularly, when a high percentage of videos are accessed frequently. In such repositories, the proposed approach reduces the incurred cost by up to 70\%.
\end{abstract}


\IEEEpeerreviewmaketitle

\section{Introduction}
Video streaming through the Internet has become a common practice. Viewers stream videos on variety of devices, from large screen TVs and desktops to tablets and smart-phones. Based on the Global Internet Phenomena Report~\cite{report}, video streaming  currently constitutes around 64\% of the U.S. Internet traffic and it is rocketing to 80\% of the whole Internet traffic by 2019\cite{forecast}. 

Video contents, either in the form of Video On Demand (VOD) (\eg YouTube\footnote{https://www.youtube.com} or Netflix\footnote{https://www.netflix.com}) or live-streaming (\eg Livestream\footnote{https://livestreams.com}), need to be converted (i.e., transcoded) based on the device characteristics of viewers. That is, the original video has to be converted to a supported characteristics (e.g., frame rate, resolution, and network bandwidth) of the viewers' devices~\cite{ahmad}.

Video transcoding is a compute-intensive and time-consuming process. To overcome the computational demand of transcoding, Video Stream Providers (VSP) extensively utilize cloud computing services~\cite{xiangbo}. To make the VOD streams readily available for viewers, VSPs commonly carry out the transcoding operation of VODs in an off-line manner. That is, they transcode and store multiple formats of the same video in the cloud to satisfy the requirements of viewers with heterogeneous display devices. We call this approach \emph{pre-transcoding}. In practice, VSPs such as Netflix have to pre-transcode 40 to 50 different formats of a single video~\cite{netflix} and store them in their repositories. This imposes a significant cost overhead to VSPs~\cite{livlsc,li1}.

Recent studies show that accessing videos of a VSP follows a long tail distribution~\cite{sharma}. That is, there are few videos that are accessed very frequently  while there is a huge portion of videos that are rarely accessed. Thus, research works have been undertaken (\eg~\cite{xiangbo,li1}) to alleviate the cost overhead of pre-transcoding by transcoding rarely-accessed videos in an on-demand (\ie lazy) manner. In this manner, one or few formats of a video is stored and transcoding is performed on-the-spot upon request to access a format of video that is not already pre-transcoded. This has become feasible with the enormous computational capacity clouds offer. We term the lazy transcoding of videos as \emph{re-transcoding} and storing of videos as \emph{pre-transcoding}. 

Re-transcoding induces the computational cost of VSPs which is generally more expensive than the storage cost~\cite{amazon}, this is because computation power are charged in on hourly basis in cloud. Therefore, the re-transcoding approach would be beneficial to VSPs, only if it is applied on the rarely accessed videos. Conversely, if it is applied on \emph{frequently accessed videos} (FAVs), it increases the cost overhead  significantly as we have to pay for every time the video stream is transcoded. 

 The challenge is how to achieve partial pre-transcoding? That is, which parts of the video should be re-transcoded and which part pre-transcoded?

To address these challenges, in this paper, we propose  a method to perform pre-transcoding on a portion of the video. 

%
%


In summary, the contributions of this paper are: \textbf{(A)} Proposing method to reduce the incurred cost of using cloud services through pre-transcoding, re-transcoding, or partially-pretranscoding videos in the repository. \textbf{(B)} Analyzing the impact of our proposed method when the rate of access to video streams in a repository varies.
   

Experimental results demonstrate that our proposed partial-transcoding method can reduce the cost overhead of VSPs significantly. The research outcome of this paper can help VSPs to reduce their cost overhead without losing Quality of Service (QoS) demanded by their viewers.

The rest of the paper is organized as follows: section II provides some background on video streaming and transcoding. Section III presents the partial pre-transcoding method. Experiment setup is detailed in section IV. Experimental results are discussed in section V, related works are presented in section VI and finally section VII concludes the paper.

\section{Background}
\subsection{Video Stream Structure}
\begin{figure}
    \includegraphics[width=8cm, height=4cm]{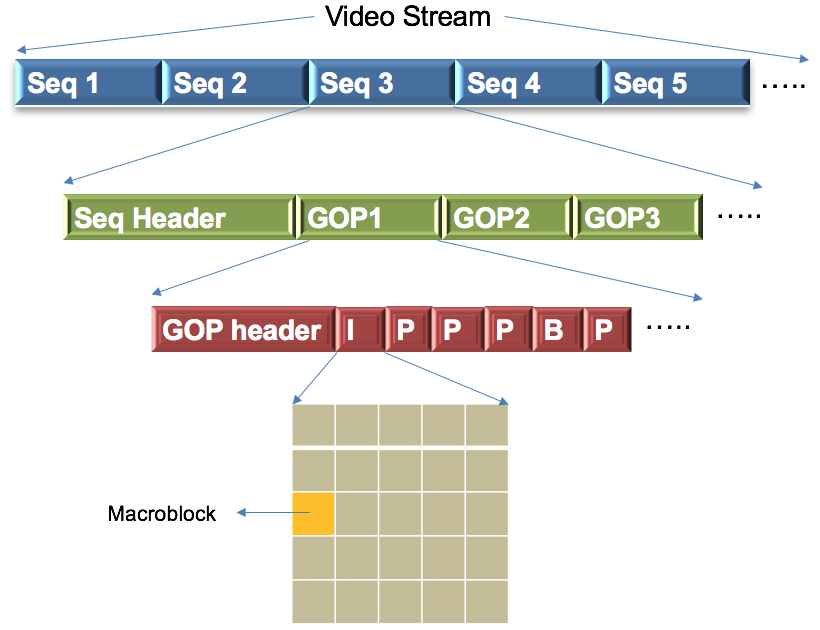}
    \caption{Video Stream Structure}
    \label {video-stream-structure}
\end{figure}

A video stream is a set of sequences as shown in Fig.~\ref{video-stream-structure}. Each sequence is built of several \emph{Group Of Pictures (GOPs)}. The first block of a sequence is called a sequence header that contains some meta-data about that sequence. Also, each GOP is constructed of a GOP header followed by several frame types, starting with I (intra) frame, followed by a P (predicted) and B (bi-directional) frames. Each frame is further comprised of slices that are formed from macroblocks (MB)~\cite{jokhio1}. 

As each GOP can be processed independently, video transcoding operation is commonly achieved at the GOP level~\cite{jokhio1}. That is, each GOP is considered as a unit for processing or pre-transcoding.  
\subsection{Cloud Services for Video Stream Transcoding}
Cloud providers offer different services in an on-demand manner and charge their users in a pay-as-you-go manner. A cloud-based video streaming system utilizes different cloud services as follows:
\begin{itemize}
\item Computational Services: Computational services in clouds are generally provided through Virtual Machines (VM) and users are charged in an hourly basis.
\item Storage Services: Users are charged for storage services based on the volume of their data stored on the cloud usually on a monthly basis. 
\item Content Delivery Network (CDN) services: CDN is a technology that reduces the delay to access different static content types, including video streams, through the Internet. CDN technology replicates the content (\eg video content) in different geographical areas to minimize the network travel time of content to users~\cite{saro,vakali}. 
\end{itemize}

Amazon Web Services\footnote{https://aws.amazon.com/} is a major cloud provider and offers all the foregoing cloud services with a high reliability. Although this study is independent of AWS technology and can work on any of cloud provider, we consider AWS services, charging model, and costs for our evaluations. 

Amazon Elastic Compute Cloud (Amazon EC2) provides computational services in form of VMs. It offers various types of VMs to cover different computational demands. General purpose \texttt{t2-small} VM type is the most common service used for different type of processing and we utilize this VM type for our evaluations as well. The hourly cost of \texttt{t2-small} VM is  \$0.026. Amazon Simple Storage Service (Amazon S3) is the storage service of Amazon cloud. Amazon S3 costs \$0.03 for each Gigabyte of stored data in a month.
Amazon also offers the CDN service (called CloudFront\footnote{https://aws.amazon.com/cloudfront/}). It delivers the content to users (\ie viewers) through a worldwide network of data centers with minimum delay~\cite{cloudfront}. Amazon charges \$0.085 for each Gigabyte of data uses CloudFront for the first 10TB per month.

It is worth noting that CDN services are required, in addition to storage services to perform pre-transcoded cloud-based video streaming. However, the CDN cost is not applied when we provide on-demand video transcoding service.


\section{Proposed Partial Pre-Transcoding Method}
\subsection{Video Streams Access Pattern}
As re-transcoding cost is incurred every time a video stream is transcoded, it is crucial to know the access pattern to videos.  Previous studies show that accessing video streams follow a long tail distribution. Other observations (\eg~\cite{miranda}) reveal that, within each video stream, the beginning segments (GOPs) are watched more frequently than the rest of the video streams. 

Miranda \etal~\cite{miranda} show that the distribution of the views within a video stream follows a long tail distribution. More specifically, they show that the distribution of accesses to GOPs of a video stream can be expressed by Power-law~\cite{newman} model. Therefore, if a video stream is accessed  $V$ times, then we can estimate the number of times each GOP is viewed. Let $GOP_{i}$ as the $i$th GOP in a video stream, then the number of views for $GOP_{i}$, denoted $P_i$, is estimated based on Equation~\ref{gop-predict}. In this equation, $\alpha$ is a constant with value 0.1.

\begin{equation}
 P_{i}=\dfrac{V}{ GOP_{i}^{\alpha} }
\label{gop-predict}
\end{equation}

\subsection{Storage Cost of Video Streams}
The storage cost of a video stream depends on the video size, and the cloud storage unit price. The equation of    cloud storage cost for video $v$ is defined as:
 \begin{equation}
C_{S}=\dfrac{S_{v} \cdot P_{S}}{2^{10}}
  \label{storage}
\end{equation} where $S_{v}$ is the size of video $v$ in MB, $P_{S}$ is the storage unit price in dollar per GB, the term $2^{10}$ is used to convert the video size unit from MB to GB. Equation~\ref{storage}, however, might be extended to compute the storage of cost of each GOP in the video stream. The equation of storage cost for $GOP_i$ is $C_{S_{i}}=\dfrac{P_{S} \cdot S_{GOP_{i}} }{2^{10}}$, where $S_{GOP_{i}}$ is the size of $GOP_i$.
\subsection{Transcoding Cost of Video Streams}
The transcoding cost is the cost of using virtual machines (VMs) that depends on the time span of utilizing VMs and the type of VM utilized for transcoding. Let $P_T$ the cost of using VM for an hour, and $\tau_{v}$ the estimated transcoding time of video $v$ in seconds. Then, the cost of transcoding of video $v$, denoted $C_{T}$, is obtained using Equation~\ref{trans}.
\begin{equation}
C_{T}= \dfrac{P_{T} \cdot \tau_{v}}{3600}
\label{trans}
\end{equation} 

It is noteworthy that Equation~\ref{trans} determines the cost for one time re-transcoding of a video. However, if a video is re-transcoded  $V$ times, the total cost would be $V\cdotp C_T$.

Estimation of transcoding execution time ($\tau_{v}$ in Equation~\ref{trans}) can be obtained based on historic transcoding execution times of the video in the past\footnote{As we are dealing with VOD, we expect that videos have been watched, thus, transcoded before. However, this is not the case when we deal with live stream videos. In that case such historic execution information is not available.}. In particular, the estimated transcoding time of video $v$ is the sum of transcoding time of all GOPs in that video, \ie $\tau_{v}=\sum_{i=1}^{m}\tau_{i}$, where $\tau_{i}$ is the estimated transcoding time of $GOP_i$ and $m$ is the total number of GOPs in video $v$.

Accordingly, Equation~\ref{trans} can be extended to compute the transcoding cost of each GOP in the video stream. The equation of transcoding cost for $GOP_i$ is represented by $C_{T_{i}}=\dfrac{P_{T} \cdot \tau_{i}}{3600}$, where $\tau_{i}$ is the estimated transcoding time of $GOP_i$.

\subsection{Algorithm \ref{al1}}

The goal of this algorithm is to minimize the incurred cost of using cloud services for VOD transcoding. For that goal, the algorithm is executed for each video in the VSP repository periodically (\eg monthly). The pseudo code for this algorithm is presented in Algorithm~\ref{al1}. It receives the video size, estimated video transcoding time, storage unit  price, transcoding unit price, and number of views of the video in the last time period (\eg last month) as input values. The output of the algorithm is whether to pre-transcode the whole video or re-transcode it upon request.

 The algorithm calculates the cost of storage and transcoding according to what was discussed in Equation~\ref{storage} and \ref{trans} ( see steps 1 and 2 Algorithm~\ref{al1}). We calculate the ratio of video storage cost to video transcoding cost in step 6 (called \emph{cost ratio}). When the ratio is less than or equal 1, it means that the cost of pre-transcoding is less than or equal the re-transcoding cost, thus,  we store that video format, otherwise, we call Algorithm~\ref{al2} to decide about partial pre-transcoding of the video. It is worth noting that for a new video  $V=0$, thus, we postpone its transcoding to when it is requested.

\subsection{Algorithm \ref{al2}}
\begin{figure}
 \includegraphics[width=8cm, height=4cm]{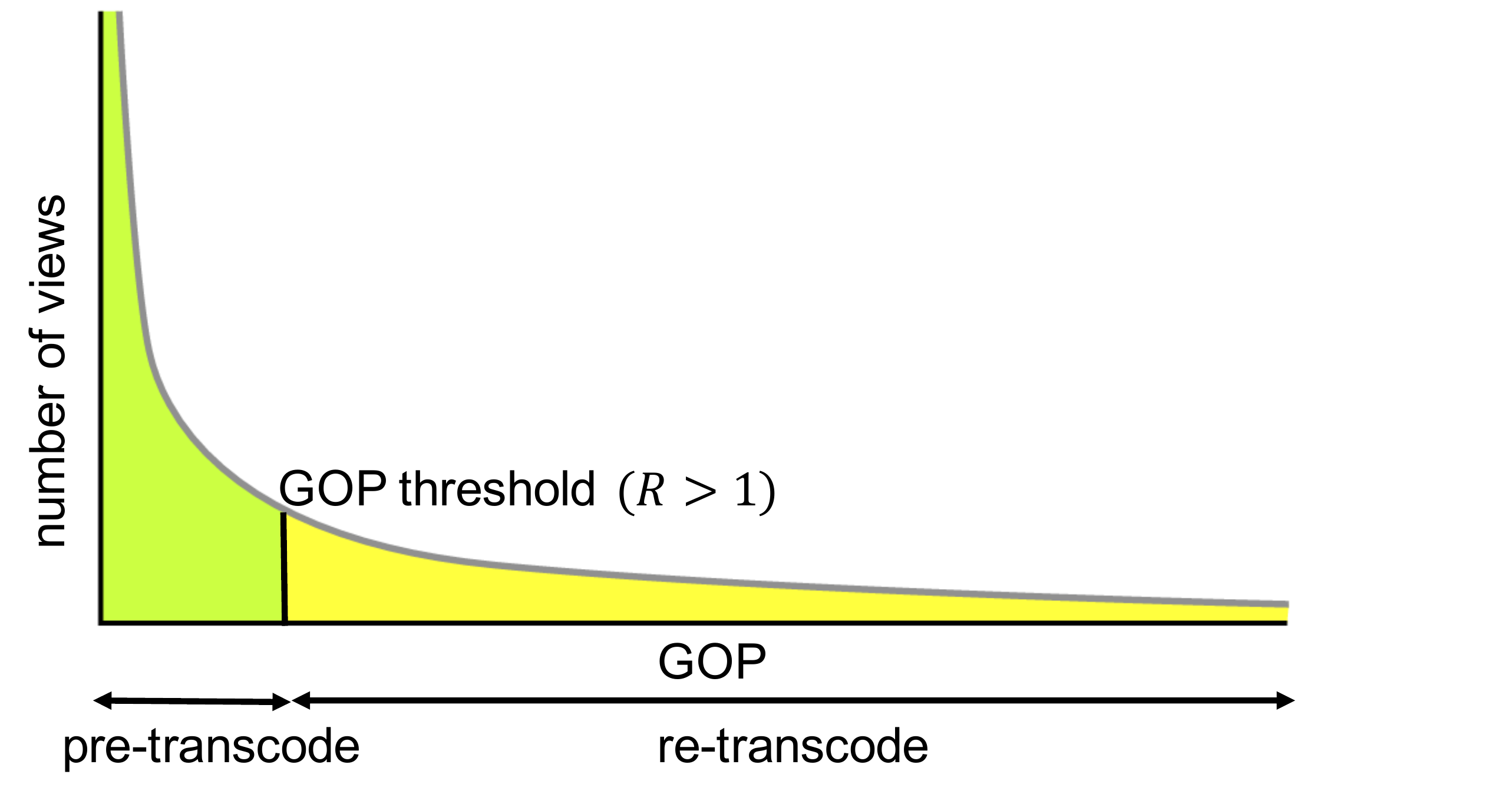}
    \caption{
    Long tail distribution of GOPs views in video stream and  partial pre-transcoding a video stream based on GOP threshold}
    \label {Video-partial4}
\end{figure}

 When the cost ratio for a video is greater than 1, Algorithm~\ref{al2} is called to possibly pre-transcode a portion of the video (termed partially pre-transcoded video). It is proven that the access pattern of GOPs within a video stream follows a long tail distribution~\cite{miranda}. That is, the first GOPs of a video stream are watched more often than the rest. Thus, the cost ratio of GOPs constantly increases for the later GOPs in the video stream. Accordingly, it is possible to find a boundary GOP so that all GOPs before it need to be pre-transcoded and all GOPs after that re-transcoded. We call this dividing point as GOP threshold ($GOP_{th}$). Formally, $GOP_{th}$ is defined as the first GOP of a video stream that has its cost ratio greater than one.
 
 Algorithm~\ref{al2} receives GOPs of a video stream, GOP size, estimated transcoding time of GOP, storage unit price, transcoding unit price and number of requests for the video stream in the last time period as input values. 
 
 Algorithm~\ref{al2} searches for the GOP threshold  by calculating the cost ratio of each GOP sequentially, starting from the first GOP in the video stream (step 1 to 5 in Algorithm~\ref{al2}). After finding the GOP threshold, Algorithm 2 pre-transcodes the GOPs before GOPthreshold and re-transcodes the GOPs after it as shown in Algorimth~\ref{al2}
 
 Fig.~\ref{Video-partial4} illustrates how Algorithm~\ref{al2} functions based on the long tail access pattern to GOPs within a video stream. Once we find the GOP threshold, all the GOPs with cost ratio less than 1 (\ie $R_i<1$) are pre-transcoded and the rest of GOP are re-transcoded.


{\SetAlgoNoLine%
\begin{algorithm}
    \SetKwInOut{Input}{Input}
    \SetKwInOut{Output}{Output}

    \Input{\\Size of video $v$: $S_{v}$\\
  Transcoding time of video $v$: $\tau_{v}$\\
  Transcoding unit price: $P_{T}$\\
  Storage unit  price: $P_{S}$\\
  Number of views in the last time period: $V$ \\

  }
    \Output{fully pre-transcoding or  fully re-transcode a video }
  
 Calculate cost of transcoding of  $v$:  $C_{T}\leftarrow\dfrac{P_{T} \cdot \tau_{v}}{3600}$  \\
    Calculate cost of storage of  $v$: $C_{S}\leftarrow\dfrac{S_{v} \cdot P_{S}}{2^{10}}$ \\
    \If{$V= 0$}
    {
    Re-transcode the whole video $v$
    }
    \eIf{$\dfrac{C_{S}}{V \cdot C_{T}}\leq 1$}
      {
       pre-transcode video $v$  
      }
      {
        Perform partial pre-transcoding using Algorithm ~\ref{al2}
      }
    \caption{ Fully pre-transcode or fully re-transcode a video}
    \label{al1}
\end{algorithm}

\begin{algorithm}
    \SetKwInOut{Input}{Input}
    \SetKwInOut{Output}{Output}

    \Input{\\ video $v$ with $m$ $GOP$\\
    Size of $GOP_{i}$: $S_{i}$ \\
    
  Transcoding time of video $GOP_{i}$: $\tau_{i}$\\
  Transcoding unit price: $P_{T}$\\
  Storage unit price: $P_{S}$\\
  Views number in the last time period : $V$\\
 
  }
    \Output{Partial pre-transcoding a video }
    For each $GOP_i$ in video stream $v$\\
     
  \Indp    
  Estimated number of views for $GOP_i$: $P_{i}\leftarrow V \cdot GOP_{i}^{-\alpha} $ \\

    Calculate cost of transcoding of $GOP_{i}$: $C_{T_{i}}\leftarrow\dfrac{P_{T} \cdot \tau_{i}}{3600}$ \\
    Calculate cost of storage of $GOP_{i}$:  $C_{S_{i}}\leftarrow\dfrac{S_{i} \cdot P_{S}}{2^{10}}$\\
    Calculate cost ratio of $GOP_i$: $R_i\leftarrow\dfrac{C_{S_{i}}}{P_{i} \cdot C_{T_{i}}}$\\
       
     \eIf{$R_i>1$}
    {$GOP_{th}\leftarrow GOP_i$\\
    Break;}
    {Continue;}
    \Indm
      Pre-transcode $ GOP_1$ to $GOP_{th-1}$ \\
       Re-transcode $GOP_{th}$ to $GOP_m$
             
    \caption{Partial pre-transcoding a video}
    \label{al2}
    
\end{algorithm}
}

\section{Experiment Setup }

\subsection{Videos Synthesis}
VSPs have large video streams repository. However, we do not have access to such repositories.
Downloading a large quantity of videos and then transcoding them is a long and costly process. Therefore, we simulate large repositories by synthesizing videos.

\begin{figure}
    \centering
    \begin{subfigure}{0.25\textwidth}
        \includegraphics[width=\textwidth]{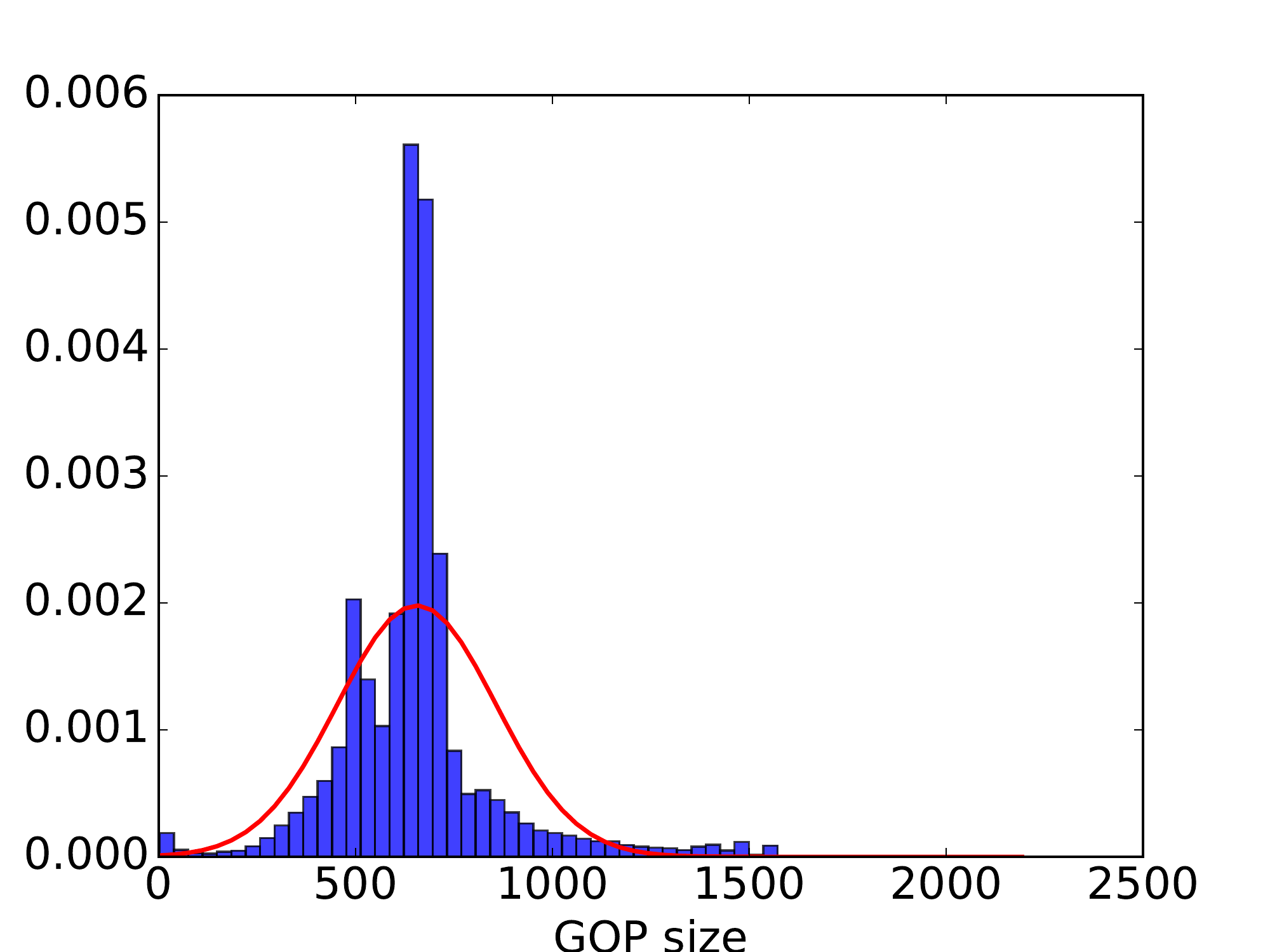}
        \caption{GOP size histogram}
        \label{fig:gopsizepdf}
    \end{subfigure}%
    \begin{subfigure}{0.25\textwidth}
        \includegraphics[width=\textwidth]{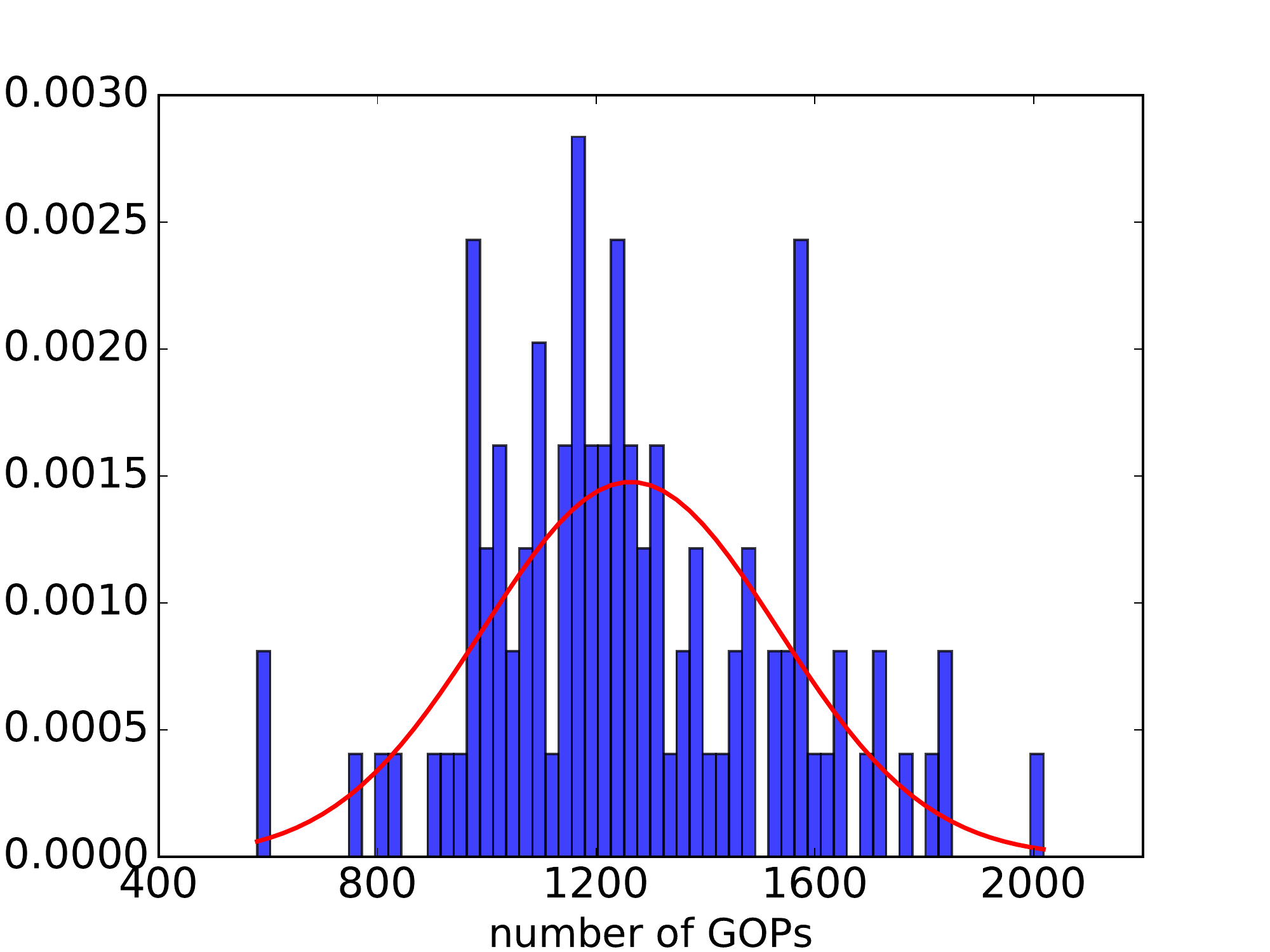}
        \caption{GOPs of video histogram}
        \label{fig:goppdf}
    \end{subfigure}%
    \caption{ distribution of GOP size  and numbers of GOPs in videos }
\end{figure}

\begin{figure}
\begin{tikzpicture}
    \pgfplotsset{width=8cm,
        compat=1.3,
        legend style={font=\footnotesize}}
    \begin{axis}[
    xlabel={GOP size (Mb)},
    ylabel={Transcoding time (s)},
    legend cell align=left,
    xmin=0,
    ymin=0,
    legend pos=north west]
    \addplot[only marks] table[row sep=\\]{
        X Y\\
      4.178676	7.722\\
3.899152	10.428\\
3.943378	9.575\\
3.931327	14.009\\
3.88854		12.761\\
2.985066	6.351\\
3.967941	9.396\\
3.921761	8.092\\
3.083184	7.046\\
2.75535		11.07\\
0.7525		2.882\\
1.75457		5.347\\
1.129115	7.317\\
6.420116	15.752\\
2.18809		14.872\\
1.07449		6.064\\
2.623932	11.974\\
0.307064	4.165\\
0.525391	2.828\\
1.240661	6.535\\
0.235102	3.149\\
0.168959	1.632\\
0.653103	1.738\\
2.405104	8.517\\
1.248916	12.731\\
0.322873	1.443\\
0.701924	5.795\\
0.005962	0.145\\
0.692842	1.727\\
1.417753	8.553\\
1.087924	5.453\\
0.965073	3.455\\
0.682937	3.783\\
0.527184	3.66\\
1.667019	9.779\\
1.971144	11.242\\
0.235808	2.142\\
1.411644	6.267\\
0.426714	3.595\\
0.309296	1.197\\
0.257945	0.839\\
0.571392	2.823\\
0.2725		1.879\\
0.305556	1.122\\
0.846596	5.336\\
0.606374	1.709\\
0.325119	2.701\\
3.039612	9.443\\
0.106311	2.845\\
0.86991		2.133\\
0.229217	1.803\\
0.580391	1.602\\
0.330577	3.655\\
0.698475	4.279\\
0.143448	1.464\\
0.386823	1.161\\
0.198863	1.314\\
1.079178	3.053\\
1.012956	3.494\\
1.086477	7.936\\
0.519122	2.775\\
0.523607	1.62\\
1.03405		4.127\\
0.191773	2.77\\
1.388117	8.965\\
0.779097	1.74\\
2.07458		10.8\\
0.423985	0.198\\
0.61006		1.973\\
0.343114	2.44\\
0.231074	1.908\\
0.371014	1.446\\
0.814905	2.928\\
0.277186	2.226\\
0.383462	1.732\\
0.329132	2.172\\
0.833752	4.129\\
2.09954		8.096\\
0.539405	2.063\\
0.654846	2.303\\
0.473374	1.885\\
0.996252	3.977\\
0.454088	6.375\\
0.968391	3.093\\
0.218652	2.855\\
0.434692	3.612\\
2.891087	7.311\\
0.837273	3.811\\
0.737742	3.975\\
0.797615	2.41\\
0.958251	4.306\\
0.710126	3.562\\
1.025744	4.003\\
0.219435	2.919\\
2.945822	14.068\\
0.664582	3.537\\
0.516766	3.401\\
3.774517	14.645\\
2.621502	10.408\\
4.303886	14.603\\
2.14266		9.242\\
0.453665	1.619\\
0.382504	2.453\\
0.459263	1.604\\
0.470892	2.186\\
0.72721		7.166\\
0.529861	1.429\\
0.171856	1.583\\
0.194693	1.107\\
0.436528	1.862\\
0.269624	0.865\\
0.682401	2.861\\
0.585747	3.892\\
0.448302	2.368\\
0.602288	1.908\\
0.633351	4.556\\
0.472549	1.783\\
0.409071	3.241\\
1.405469	1.889\\
2.273343	4.877\\
0.654088	3.72\\
1.022323	1.429\\
0.325187	2.898\\
0.499621	1.964\\
2.880156	4.795\\
0.466623	3.894\\
1.291018	5.954\\
0.359753	1.341\\
0.884227	0.745\\
0.845712	2.089\\
0.79096		7.694\\
0.914541	6.569\\
3.089503	11.374\\
1.695496	9.963\\
0.187188	2.539\\
0.983811	7.584\\
0.124924	1.913\\
0.232798	2.179\\
0.327905	1.55\\
0.317465	3.096\\
0.780119	5.199\\
2.050206	13.936\\
2.791362	17.604\\
3.219139	20.862\\
3.383975	19.963\\
5.741488	14.881\\
6.491819	12.872\\
6.219561	14.614\\
2.423557	14.438\\
0.891129	8.829\\
0.914171	7.284\\
0.69056		4.318\\
3.234574	13.741\\
0.951665	3.165\\
2.01185		6.438\\
0.998136	7.93\\
6.644439	18.693\\
2.77579		16.901\\
1.533786	7.474\\
2.807641	12.731\\
0.583437	0.143\\
0.287397	2.824\\
0.49798		3.25\\
1.255763	7.271\\
0.177287	2.991\\
0.149504	1.619\\
0.660657	1.77\\
2.854946	8.352\\
1.450513	11.577\\
0.342982	1.255\\
0.643656	5.907\\
0.49728		3.513\\
1.381017	8.941\\
1.012388	5.761\\
1.054952	3.849\\
0.642743	4.225\\
0.55854		3.928\\
1.815247	10.402\\
1.95212		11.382\\
0.210778	2.173\\
1.339076	6.444\\
0.204333	3.015\\
0.610726	0.498\\
0.254033	0.9\\
0.551058	2.971\\
0.235067	1.877\\
0.290101	1.328\\
0.771734	5.824\\
0.526757	1.894\\
0.515342	3.475\\
2.602418	11.557\\
0.085168	3.151\\
0.868362	2.052\\
0.264047	1.877\\
0.569402	1.713\\
0.32306		4.183\\
0.682008	4.6\\
0.147494	1.704\\
0.368608	1.152\\
0.197569	1.595\\
1.077269	3.184\\
1.070046	3.646\\
1.122351	8.802\\
0.391208	3.067\\
0.445168	1.316\\
0.927374	4.313\\
0.15976		2.847\\
1.520942	10.941\\
0.656027	2.117\\
2.471742	13.547\\
0.204229	1.782\\
0.609737	2.08\\
0.310522	2.558\\
0.219383	1.918\\
0.34552		1.536\\
0.764382	3.07\\
0.305451	2.199\\
0.357993	2.039\\
0.28775		2.189\\
0.864573	4.614\\
1.880662	8.809\\
0.52929		2.301\\
0.595715	2.553\\
0.393251	1.949\\
0.863181	4.062\\
0.41071		6.862\\
0.99823		3.284\\
0.218477	2.523\\
3.216722	11.381\\
0.721097	3.929\\
0.785707	4.136\\
0.792496	2.606\\
0.974937	4.355\\
0.748311	3.83\\
1.062749	4.165\\
0.221227	2.999\\
2.841728	14.13\\
0.726273	3.691\\
0.413349	3.759\\
2.386438	7.266\\
4.255978	17.383\\
4.188468	14.842\\
2.402441	9.627\\
0.377929	1.685\\
0.344836	2.321\\
0.41787		1.578\\
0.468973	2.196\\
0.713922	7.401\\
0.437848	1.49\\
0.135652	1.497\\
0.202143	1.134\\
0.44318		2.08\\
0.995009	5.301\\
0.483138	3.789\\
0.476255	2.102\\
0.577433	1.934\\
0.897766	4.64\\
0.336884	1.879\\
0.39692		3.236\\
1.390246	1.819\\
1.849219	4.769\\
0.642075	3.826\\
0.964541	1.598\\
0.37586		3.308\\
0.495524	2.508\\
2.720884	5.389\\
0.504648	4.455\\
1.25113		6.587\\
0.319503	1.358\\
0.851909	0.756\\
0.86409		2.142\\
0.918454	8.599\\
0.943907	6.602\\
3.247761	11.157\\
2.284932	10.223\\
0.18423		2.635\\
1.196954	7.687\\
0.186771	2.46\\
0.200448	2.904\\
0.351297	1.416\\
0.308545	3.245\\
1.836131	12.527\\
2.578704	19.239\\
3.075134	20.168\\
3.09592		19.748\\
3.838922	17.142\\
4.262899	13.001\\
5.187993	14.574\\
3.083568	16.132\\
1.463361	13.965\\
0.842132	7.359\\
0.673031	4.597\\
0.137231	1.492\\
0.455165	0.813\\
1.177364	7.592\\
0.375538	1.05\\
0.843325	4.655\\
1.326349	3.035\\
0.364347	2.154\\
0.268786	1.291\\
0.457276	1.229\\
0.055503	0.782\\
0.256329	0.1\\
0.325894	1.296\\
0.268261	1.61\\
0.41499		1.204\\
0.699243	2.727\\
1.250638	3.523\\
0.930825	3.971\\
1.475622	4.773\\
0.240949	1.335\\
1.094136	3.799\\
0.457841	3.827\\
0.730164	1.079\\
0.141969	1.363\\
0.585104	2.663\\
1.109746	1.767\\
0.631111	2.808\\
1.221043	2.242\\
0.451512	2.712\\
0.386564	2.179\\
0.131462	1.084\\
0.851757	1.746\\
0.562499	2.228\\
0.121405	1.676\\
0.427288	1.83\\
0.460556	2.022\\
0.146059	1.679\\
0.034276	0.487\\
0.587572	0.192\\
0.61183		1.737\\
0.106897	2.512\\
0.332454	1.235\\
0.327102	0.995\\
0.437725	2.943\\
0.576592	3.298\\
0.466881	3.563\\
1.118472	2.254\\
0.525937	2.608\\
0.79291		2.254\\
0.659349	3.357\\
0.8102		1.053\\
0.671828	2.379\\
0.361761	3.475\\
0.565995	0.802\\
0.276157	1.638\\
0.848709	2.035\\
0.620053	1.76\\
0.952145	2.547\\
0.368201	1.843\\
0.668109	3.005\\
0.871042	2.392\\
0.964527	1.913\\
0.424161	2.065\\
0.919755	2.241\\
0.668033	2.303\\
0.397926	2.124\\
0.349954	1.611\\
0.442978	1.72\\
0.207251	0.443\\
0.5593		0.978\\
0.46376		3.438\\
0.446078	3.23\\
0.531405	3.384\\
0.454697	3.73\\
0.247309	2.215\\
0.358423	2.191\\
0.258195	1.451\\
0.539075	3.891\\
0.432662	3.714\\
0.464885	3.32\\
0.316642	3.12\\
0.295292	2.171\\
0.174789	2.256\\
0.127743	1.69\\
0.067966	0.828\\
0.164133	1.734\\
0.169318	1.748\\
0.217121	1.915\\
0.231042	2.175\\
0.308859	2.247\\
0.333274	2.636\\
0.231739	3.027\\
0.139301	0.826\\
0.215682	2.051\\
0.153764	1.882\\
0.060221	1.492\\
0.071315	1.443\\
0.424254	1.612\\
0.582353	4.288\\
1.018014	4.887\\
0.479808	2.621\\
1.174496	4.284\\
0.101063	1.54\\
0.101748	0.691\\
0.055798	0.775\\
0.093361	0.808\\
0.255016	1.014\\
0.061761	0.953\\
0.053304	0.452\\
0.159349	1.178\\
0.185292	1.506\\
0.129017	1.451\\
0.0886		1.256\\
0.052344	1.023\\
0.211098	1.119\\
0.875946	2.39\\
0.456097	3.419\\
0.381189	4.565\\
0.275241	2.479\\
0.159211	2.096\\
0.149617	1.937\\
0.149561	1.868\\
0.226508	1.675\\
0.329765	1.957\\
0.122389	1.009\\
0.324116	1.891\\
0.376919	3.041\\
0.316309	4.548\\
0.260456	3.328\\
0.417955	3.328\\
0.409756	5.641\\
0.347501	4.375\\
0.185232	2.326\\
0.315772	3.742\\
0.274141	3.661\\
0.256076	3.362\\
0.213985	2.816\\
0.198398	2.33\\
0.25527		2.128\\
0.216398	2.448\\
0.121182	1.162\\
0.720187	2.93\\
0.647845	4.811\\
0.646509	4.282\\
0.436049	4.392\\
0.000288	0.756\\
0.203487	1.88\\
1.015559	0.72\\
0.889221	3.859\\
0.491204	2.541\\
0.63234		2.699\\
0.541576	2.365\\
0.354727	1.38\\
0.474095	1.785\\
0.250803	1.979\\
0.673341	3.411\\
0.628855	2.293\\
0.225852	2.598\\
0.380692	0.858\\
0.492996	1.905\\
0.264938	1.201\\
1.007422	3.24\\
1.168786	5.522\\
0.612628	2.578\\
0.878793	5.275\\
0.114312	0.783\\
0.926268	0.22\\
0.376676	1.168\\
0.576182	2.462\\
0.169886	2.809\\
0.300662	1.145\\
0.954408	1.303\\
0.627746	3.035\\
0.70329		2.275\\
0.382945	1.777\\
0.162393	2.026\\
0.320834	2.44\\
0.668309	3.194\\
0.772603	2.16\\
0.456091	3.283\\
1.07013		4.216\\
1.749542	7.155\\
0.486427	2.138\\
0.751509	2.734\\
0.366677	3.574\\
0.643496	1.716\\
0.775366	0.246\\
0.311321	2.603\\
0.267283	2.439\\
0.797449	4.264\\
0.657886	2.96\\
0.415861	2.805\\
0.543426	2.965\\
0.452297	1.94\\
0.620149	1.602\\
0.192103	2.406\\
0.449135	2.141\\
0.326877	1.545\\
0.749811	2.977\\
0.322546	2.472\\
0.354046	2.084\\
0.497748	5.005\\
0.104998	1.38\\
0.595122	1.242\\
0.781183	1.812\\
0.735793	4.187\\
0.853004	3\\
0.418552	2.287\\
0.82742		1.994\\
0.299448	2.227\\
0.313784	2.338\\
0.42274		1.61\\
0.501348	1.579\\
0.564264	1.126\\
0.417449	1.891\\
0.552053	2.531\\
0.530747	3.332\\
0.681495	2.438\\
0.687057	0.188\\
0.74649		1.339\\
0.376384	2.046\\
0.755948	2.629\\
0.64104		2.063\\
0.505403	1.445\\
1.399299	6.85\\
0.247867	1.499\\
0.230224	1.635\\
0.877222	3.063\\
0.105117	1.619\\
0.09609		1.476\\
0.221719	0.883\\
1.801817	9.968\\
0.011841	2.261\\
0.827845	4.83\\
0.426733	0.769\\
0.387433	4.808\\
3.055171	9.083\\
0.76146		5.304\\
0.00098		0.1\\
0.935574	0.24\\
0.789459	0.757\\
0.422027	2.399\\
0.770927	2.698\\
1.110919	1.947\\
1.074328	6.666\\
1.53774		2.678\\
1.479017	3.466\\
1.123666	4.485\\
0.92599		4.334\\
1.297341	4.024\\
1.549623	6.052\\
1.108864	2.881\\
1.193374	3.631\\
1.067844	3.211\\
0.518198	2.729\\
1.343571	3.348\\
1.204393	3.018\\
1.587163	4.436\\
3.119053	8.998\\
0.897657	4.026\\
0.4674		5.524\\
0.85974		3.073\\
1.404266	2.628\\
0.857686	1.448\\
0.927745	3.493\\
0.423697	5.29\\
3.71795		9.208\\
1.013562	7.55\\
0.916351	3.679\\
1.433096	5.415\\
1.284562	4.282\\
1.14336		4.771\\
1.041961	6.341\\
1.419648	4.486\\
0.863947	3.127\\
1.045521	7.315\\
0.4957		3.656\\
0.248571	3.191\\
0.553977	1.82\\
0.095839	1.975\\
0.194775	1.17\\
1.494827	6.349\\
1.13516		6.283\\
1.603009	7.379\\
1.632214	3.053\\
0.911846	2.573\\
0.527703	2.639\\
0.454461	1.427\\
0.293249	1.427\\
0.451601	1.629\\
0.173344	0.845\\
0.792472	1.877\\
0.086966	1.374\\
0.196223	0.066\\
0.217286	0.29\\
0.613043	0.928\\
0.754553	1.288\\
0.94617		0.921\\
0.19008		1.4\\
0.231557	1.897\\
0.134755	0.503\\
0.180727	0.661\\
0.233209	1.226\\
0.348287	1.142\\
0.001593	0.1\\
0.487917	0.113\\
0.519924	0.4\\
0.447908	2.386\\
2.422342	1.523\\
0.890386	2.182\\
0.578587	4.113\\
0.390558	0.941\\
0.655013	1.204\\
0.250455	1.251\\
0.6573		1.476\\
0.13822		1.264\\
0.289158	1.295\\
0.795728	0.168\\
0.658959	0.751\\
0.271119	1.244\\
0.312418	1.035\\
0.315582	0.958\\
0.629757	0.78\\
0.316039	0.916\\
0.447758	1.155\\
0.343717	0.519\\
0.527189	1.547\\
0.489795	0.872\\
0.528911	0.922\\
0.640295	1.663\\
0.568387	0.125\\
0.138075	0.363\\
0.616623	1.221\\
0.120954	0.611\\
1.320342	2.019\\
0.388337	1.423\\
0.830432	1.747\\
0.358069	0.968\\
0.235508	1.513\\
0.270689	1.915\\
0.548536	2.081\\
0.9118		0.859\\
0.517615	1.267\\
0.212366	1.513\\
0.92469		1.043\\
0.786187	1.294\\
0.214037	2.021\\
1.089054	3.147\\
0.258219	1.476\\
0.23902		2.223\\
0.466037	1.764\\
0.697814	4.407\\
2.037389	10.627\\
1.285648	7.088\\
1.227666	7.02\\
1.645516	8.716\\
1.068248	7.525\\
0.757438	4.861\\
1.100554	7.601\\
0.648771	6.191\\
    };
    
  \addlegendentry{data points}
    \addplot table[row sep=\\,
    y={create col/linear regression={y=Y}}] 
    {
       X        Y\\
 4.178676	7.722\\
3.899152	10.428\\
3.943378	9.575\\
3.931327	14.009\\
3.88854	12.761\\
2.985066	6.351\\
3.967941	9.396\\
3.921761	8.092\\
3.083184	7.046\\
2.75535	11.07\\
0.7525	2.882\\
1.75457	5.347\\
1.129115	7.317\\
6.420116	15.752\\
2.18809	14.872\\
1.07449	6.064\\
2.623932	11.974\\
0.307064	4.165\\
0.525391	2.828\\
1.240661	6.535\\
0.235102	3.149\\
0.168959	1.632\\
0.653103	1.738\\
2.405104	8.517\\
1.248916	12.731\\
0.322873	1.443\\
0.701924	5.795\\
0.005962	0.145\\
0.692842	1.727\\
1.417753	8.553\\
1.087924	5.453\\
0.965073	3.455\\
0.682937	3.783\\
0.527184	3.66\\
1.667019	9.779\\
1.971144	11.242\\
0.235808	2.142\\
1.411644	6.267\\
0.426714	3.595\\
0.309296	1.197\\
0.257945	0.839\\
0.571392	2.823\\
0.2725	1.879\\
0.305556	1.122\\
0.846596	5.336\\
0.606374	1.709\\
0.325119	2.701\\
3.039612	9.443\\
0.106311	2.845\\
0.86991	2.133\\
0.229217	1.803\\
0.580391	1.602\\
0.330577	3.655\\
0.698475	4.279\\
0.143448	1.464\\
0.386823	1.161\\
0.198863	1.314\\
1.079178	3.053\\
1.012956	3.494\\
1.086477	7.936\\
0.519122	2.775\\
0.523607	1.62\\
1.03405	4.127\\
0.191773	2.77\\
1.388117	8.965\\
0.779097	1.74\\
2.07458	10.8\\
0.423985	0.198\\
0.61006	1.973\\
0.343114	2.44\\
0.231074	1.908\\
0.371014	1.446\\
0.814905	2.928\\
0.277186	2.226\\
0.383462	1.732\\
0.329132	2.172\\
0.833752	4.129\\
2.09954	8.096\\
0.539405	2.063\\
0.654846	2.303\\
0.473374	1.885\\
0.996252	3.977\\
0.454088	6.375\\
0.968391	3.093\\
0.218652	2.855\\
0.434692	3.612\\
2.891087	7.311\\
0.837273	3.811\\
0.737742	3.975\\
0.797615	2.41\\
0.958251	4.306\\
0.710126	3.562\\
1.025744	4.003\\
0.219435	2.919\\
2.945822	14.068\\
0.664582	3.537\\
0.516766	3.401\\
3.774517	14.645\\
2.621502	10.408\\
4.303886	14.603\\
2.14266	9.242\\
0.453665	1.619\\
0.382504	2.453\\
0.459263	1.604\\
0.470892	2.186\\
0.72721	7.166\\
0.529861	1.429\\
0.171856	1.583\\
0.194693	1.107\\
0.436528	1.862\\
0.269624	0.865\\
0.682401	2.861\\
0.585747	3.892\\
0.448302	2.368\\
0.602288	1.908\\
0.633351	4.556\\
0.472549	1.783\\
0.409071	3.241\\
1.405469	1.889\\
2.273343	4.877\\
0.654088	3.72\\
1.022323	1.429\\
0.325187	2.898\\
0.499621	1.964\\
2.880156	4.795\\
0.466623	3.894\\
1.291018	5.954\\
0.359753	1.341\\
0.884227	0.745\\
0.845712	2.089\\
0.79096	7.694\\
0.914541	6.569\\
3.089503	11.374\\
1.695496	9.963\\
0.187188	2.539\\
0.983811	7.584\\
0.124924	1.913\\
0.232798	2.179\\
0.327905	1.55\\
0.317465	3.096\\
0.780119	5.199\\
2.050206	13.936\\
2.791362	17.604\\
3.219139	20.862\\
3.383975	19.963\\
5.741488	14.881\\
6.491819	12.872\\
6.219561	14.614\\
2.423557	14.438\\
0.891129	8.829\\
0.914171	7.284\\
0.69056	4.318\\
3.234574	13.741\\
0.951665	3.165\\
2.01185	6.438\\
0.998136	7.93\\
6.644439	18.693\\
2.77579	16.901\\
1.533786	7.474\\
2.807641	12.731\\
0.583437	0.143\\
0.287397	2.824\\
0.49798	3.25\\
1.255763	7.271\\
0.177287	2.991\\
0.149504	1.619\\
0.660657	1.77\\
2.854946	8.352\\
1.450513	11.577\\
0.342982	1.255\\
0.643656	5.907\\
0.49728	3.513\\
1.381017	8.941\\
1.012388	5.761\\
1.054952	3.849\\
0.642743	4.225\\
0.55854	3.928\\
1.815247	10.402\\
1.95212	11.382\\
0.210778	2.173\\
1.339076	6.444\\
0.204333	3.015\\
0.610726	0.498\\
0.254033	0.9\\
0.551058	2.971\\
0.235067	1.877\\
0.290101	1.328\\
0.771734	5.824\\
0.526757	1.894\\
0.515342	3.475\\
2.602418	11.557\\
0.085168	3.151\\
0.868362	2.052\\
0.264047	1.877\\
0.569402	1.713\\
0.32306	4.183\\
0.682008	4.6\\
0.147494	1.704\\
0.368608	1.152\\
0.197569	1.595\\
1.077269	3.184\\
1.070046	3.646\\
1.122351	8.802\\
0.391208	3.067\\
0.445168	1.316\\
0.927374	4.313\\
0.15976	2.847\\
1.520942	10.941\\
0.656027	2.117\\
2.471742	13.547\\
0.204229	1.782\\
0.609737	2.08\\
0.310522	2.558\\
0.219383	1.918\\
0.34552	1.536\\
0.764382	3.07\\
0.305451	2.199\\
0.357993	2.039\\
0.28775	2.189\\
0.864573	4.614\\
1.880662	8.809\\
0.52929	2.301\\
0.595715	2.553\\
0.393251	1.949\\
0.863181	4.062\\
0.41071	6.862\\
0.99823	3.284\\
0.218477	2.523\\
3.216722	11.381\\
0.721097	3.929\\
0.785707	4.136\\
0.792496	2.606\\
0.974937	4.355\\
0.748311	3.83\\
1.062749	4.165\\
0.221227	2.999\\
2.841728	14.13\\
0.726273	3.691\\
0.413349	3.759\\
2.386438	7.266\\
4.255978	17.383\\
4.188468	14.842\\
2.402441	9.627\\
0.377929	1.685\\
0.344836	2.321\\
0.41787	1.578\\
0.468973	2.196\\
0.713922	7.401\\
0.437848	1.49\\
0.135652	1.497\\
0.202143	1.134\\
0.44318	2.08\\
0.995009	5.301\\
0.483138	3.789\\
0.476255	2.102\\
0.577433	1.934\\
0.897766	4.64\\
0.336884	1.879\\
0.39692	3.236\\
1.390246	1.819\\
1.849219	4.769\\
0.642075	3.826\\
0.964541	1.598\\
0.37586	3.308\\
0.495524	2.508\\
2.720884	5.389\\
0.504648	4.455\\
1.25113	6.587\\
0.319503	1.358\\
0.851909	0.756\\
0.86409	2.142\\
0.918454	8.599\\
0.943907	6.602\\
3.247761	11.157\\
2.284932	10.223\\
0.18423	2.635\\
1.196954	7.687\\
0.186771	2.46\\
0.200448	2.904\\
0.351297	1.416\\
0.308545	3.245\\
1.836131	12.527\\
2.578704	19.239\\
3.075134	20.168\\
3.09592	19.748\\
3.838922	17.142\\
4.262899	13.001\\
5.187993	14.574\\
3.083568	16.132\\
1.463361	13.965\\
0.842132	7.359\\
0.673031	4.597\\
0.137231	1.492\\
0.455165	0.813\\
1.177364	7.592\\
0.375538	1.05\\
0.843325	4.655\\
1.326349	3.035\\
0.364347	2.154\\
0.268786	1.291\\
0.457276	1.229\\
0.055503	0.782\\
0.256329	0.1\\
0.325894	1.296\\
0.268261	1.61\\
0.41499	1.204\\
0.699243	2.727\\
1.250638	3.523\\
0.930825	3.971\\
1.475622	4.773\\
0.240949	1.335\\
1.094136	3.799\\
0.457841	3.827\\
0.730164	1.079\\
0.141969	1.363\\
0.585104	2.663\\
1.109746	1.767\\
0.631111	2.808\\
1.221043	2.242\\
0.451512	2.712\\
0.386564	2.179\\
0.131462	1.084\\
0.851757	1.746\\
0.562499	2.228\\
0.121405	1.676\\
0.427288	1.83\\
0.460556	2.022\\
0.146059	1.679\\
0.034276	0.487\\
0.587572	0.192\\
0.61183	1.737\\
0.106897	2.512\\
0.332454	1.235\\
0.327102	0.995\\
0.437725	2.943\\
0.576592	3.298\\
0.466881	3.563\\
1.118472	2.254\\
0.525937	2.608\\
0.79291	2.254\\
0.659349	3.357\\
0.8102	1.053\\
0.671828	2.379\\
0.361761	3.475\\
0.565995	0.802\\
0.276157	1.638\\
0.848709	2.035\\
0.620053	1.76\\
0.952145	2.547\\
0.368201	1.843\\
0.668109	3.005\\
0.871042	2.392\\
0.964527	1.913\\
0.424161	2.065\\
0.919755	2.241\\
0.668033	2.303\\
0.397926	2.124\\
0.349954	1.611\\
0.442978	1.72\\
0.207251	0.443\\
0.5593	0.978\\
0.46376	3.438\\
0.446078	3.23\\
0.531405	3.384\\
0.454697	3.73\\
0.247309	2.215\\
0.358423	2.191\\
0.258195	1.451\\
0.539075	3.891\\
0.432662	3.714\\
0.464885	3.32\\
0.316642	3.12\\
0.295292	2.171\\
0.174789	2.256\\
0.127743	1.69\\
0.067966	0.828\\
0.164133	1.734\\
0.169318	1.748\\
0.217121	1.915\\
0.231042	2.175\\
0.308859	2.247\\
0.333274	2.636\\
0.231739	3.027\\
0.139301	0.826\\
0.215682	2.051\\
0.153764	1.882\\
0.060221	1.492\\
0.071315	1.443\\
0.424254	1.612\\
0.582353	4.288\\
1.018014	4.887\\
0.479808	2.621\\
1.174496	4.284\\
0.101063	1.54\\
0.101748	0.691\\
0.055798	0.775\\
0.093361	0.808\\
0.255016	1.014\\
0.061761	0.953\\
0.053304	0.452\\
0.159349	1.178\\
0.185292	1.506\\
0.129017	1.451\\
0.0886	1.256\\
0.052344	1.023\\
0.211098	1.119\\
0.875946	2.39\\
0.456097	3.419\\
0.381189	4.565\\
0.275241	2.479\\
0.159211	2.096\\
0.149617	1.937\\
0.149561	1.868\\
0.226508	1.675\\
0.329765	1.957\\
0.122389	1.009\\
0.324116	1.891\\
0.376919	3.041\\
0.316309	4.548\\
0.260456	3.328\\
0.417955	3.328\\
0.409756	5.641\\
0.347501	4.375\\
0.185232	2.326\\
0.315772	3.742\\
0.274141	3.661\\
0.256076	3.362\\
0.213985	2.816\\
0.198398	2.33\\
0.25527	2.128\\
0.216398	2.448\\
0.121182	1.162\\
0.720187	2.93\\
0.647845	4.811\\
0.646509	4.282\\
0.436049	4.392\\
0.000288	0.756\\
0.203487	1.88\\
1.015559	0.72\\
0.889221	3.859\\
0.491204	2.541\\
0.63234	2.699\\
0.541576	2.365\\
0.354727	1.38\\
0.474095	1.785\\
0.250803	1.979\\
0.673341	3.411\\
0.628855	2.293\\
0.225852	2.598\\
0.380692	0.858\\
0.492996	1.905\\
0.264938	1.201\\
1.007422	3.24\\
1.168786	5.522\\
0.612628	2.578\\
0.878793	5.275\\
0.114312	0.783\\
0.926268	0.22\\
0.376676	1.168\\
0.576182	2.462\\
0.169886	2.809\\
0.300662	1.145\\
0.954408	1.303\\
0.627746	3.035\\
0.70329	2.275\\
0.382945	1.777\\
0.162393	2.026\\
0.320834	2.44\\
0.668309	3.194\\
0.772603	2.16\\
0.456091	3.283\\
1.07013	4.216\\
1.749542	7.155\\
0.486427	2.138\\
0.751509	2.734\\
0.366677	3.574\\
0.643496	1.716\\
0.775366	0.246\\
0.311321	2.603\\
0.267283	2.439\\
0.797449	4.264\\
0.657886	2.96\\
0.415861	2.805\\
0.543426	2.965\\
0.452297	1.94\\
0.620149	1.602\\
0.192103	2.406\\
0.449135	2.141\\
0.326877	1.545\\
0.749811	2.977\\
0.322546	2.472\\
0.354046	2.084\\
0.497748	5.005\\
0.104998	1.38\\
0.595122	1.242\\
0.781183	1.812\\
0.735793	4.187\\
0.853004	3\\
0.418552	2.287\\
0.82742	1.994\\
0.299448	2.227\\
0.313784	2.338\\
0.42274	1.61\\
0.501348	1.579\\
0.564264	1.126\\
0.417449	1.891\\
0.552053	2.531\\
0.530747	3.332\\
0.681495	2.438\\
0.687057	0.188\\
0.74649	1.339\\
0.376384	2.046\\
0.755948	2.629\\
0.64104	2.063\\
0.505403	1.445\\
1.399299	6.85\\
0.247867	1.499\\
0.230224	1.635\\
0.877222	3.063\\
0.105117	1.619\\
0.09609	1.476\\
0.221719	0.883\\
1.801817	9.968\\
0.011841	2.261\\
0.827845	4.83\\
0.426733	0.769\\
0.387433	4.808\\
3.055171	9.083\\
0.76146	5.304\\
0.00098	0.1\\
0.935574	0.24\\
0.789459	0.757\\
0.422027	2.399\\
0.770927	2.698\\
1.110919	1.947\\
1.074328	6.666\\
1.53774	2.678\\
1.479017	3.466\\
1.123666	4.485\\
0.92599	4.334\\
1.297341	4.024\\
1.549623	6.052\\
1.108864	2.881\\
1.193374	3.631\\
1.067844	3.211\\
0.518198	2.729\\
1.343571	3.348\\
1.204393	3.018\\
1.587163	4.436\\
3.119053	8.998\\
0.897657	4.026\\
0.4674	5.524\\
0.85974	3.073\\
1.404266	2.628\\
0.857686	1.448\\
0.927745	3.493\\
0.423697	5.29\\
3.71795	9.208\\
1.013562	7.55\\
0.916351	3.679\\
1.433096	5.415\\
1.284562	4.282\\
1.14336	4.771\\
1.041961	6.341\\
1.419648	4.486\\
0.863947	3.127\\
1.045521	7.315\\
0.4957	3.656\\
0.248571	3.191\\
0.553977	1.82\\
0.095839	1.975\\
0.194775	1.17\\
1.494827	6.349\\
1.13516	6.283\\
1.603009	7.379\\
1.632214	3.053\\
0.911846	2.573\\
0.527703	2.639\\
0.454461	1.427\\
0.293249	1.427\\
0.451601	1.629\\
0.173344	0.845\\
0.792472	1.877\\
0.086966	1.374\\
0.196223	0.066\\
0.217286	0.29\\
0.613043	0.928\\
0.754553	1.288\\
0.94617	0.921\\
0.19008	1.4\\
0.231557	1.897\\
0.134755	0.503\\
0.180727	0.661\\
0.233209	1.226\\
0.348287	1.142\\
0.001593	0.1\\
0.487917	0.113\\
0.519924	0.4\\
0.447908	2.386\\
2.422342	1.523\\
0.890386	2.182\\
0.578587	4.113\\
0.390558	0.941\\
0.655013	1.204\\
0.250455	1.251\\
0.6573	1.476\\
0.13822	1.264\\
0.289158	1.295\\
0.795728	0.168\\
0.658959	0.751\\
0.271119	1.244\\
0.312418	1.035\\
0.315582	0.958\\
0.629757	0.78\\
0.316039	0.916\\
0.447758	1.155\\
0.343717	0.519\\
0.527189	1.547\\
0.489795	0.872\\
0.528911	0.922\\
0.640295	1.663\\
0.568387	0.125\\
0.138075	0.363\\
0.616623	1.221\\
0.120954	0.611\\
1.320342	2.019\\
0.388337	1.423\\
0.830432	1.747\\
0.358069	0.968\\
0.235508	1.513\\
0.270689	1.915\\
0.548536	2.081\\
0.9118	0.859\\
0.517615	1.267\\
0.212366	1.513\\
0.92469	1.043\\
0.786187	1.294\\
0.214037	2.021\\
1.089054	3.147\\
0.258219	1.476\\
0.23902	2.223\\
0.466037	1.764\\
0.697814	4.407\\
2.037389	10.627\\
1.285648	7.088\\
1.227666	7.02\\
1.645516	8.716\\
1.068248	7.525\\
0.757438	4.861\\
1.100554	7.601\\
0.648771	6.191\\
    };
    
    \addlegendentry{%
            $\pgfmathprintnumber{\pgfplotstableregressiona} \cdot x
        \pgfmathprintnumber[print sign]{\pgfplotstableregressionb}$  ($R^2=0.67$)  } %
         
    \end{axis}
     
       \end{tikzpicture}
       
    \caption{Linear regression analysis of GOP size and transcoding time}
    \label{fig:regres}
\end{figure}
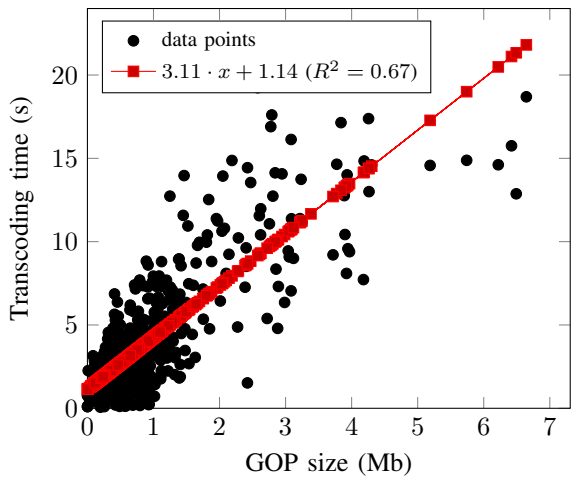

To accurately synthesize videos, we need to know the distribution of  characteristics  of  videos. Specifically, GOP size, GOP transcoding time, and number of GOPs for each video are required. We synthesize videos based on our repository that includes 103 videos of a wide range of categories including movies, animation, sports, documentary, news, and music. the size of the videos are between 400MB and 2.5GB. In order to extract the characteristics of those videos, we transcode  them by implementing the method in~\cite{xiangbo}  on a machine with 2.8 GHz Intel Core i5, 8GB RAM,  and 1TB Hard Disk. We analyze the characteristic data extracted from videos and
 observe that GOP size and number of GOPs follow a Gaussian curve, shown in Fig.~\ref{fig:gopsizepdf}  and \ref{fig:goppdf}. The parameters of the Gaussian distributions are given in table~\ref{statistics}. Additionally, Li \etal  \cite{tech}  proved through a regression analysis that there is a linear relationship between GOP size and its transcoding time. By implementing the same method we derive a linear equation between GOP sizes and their corresponding transcoding times for all downloaded videos and the plot of the analysis is shown in Fig.~\ref{fig:regres}. 
 
Having the statistics of GOP size, number of GOPs,  and  the linear equation for GOP transcoding times, we generate 50,000 videos with 112.98 TB size for our synthesized repository.
 \begin{table}
\centering
\begin{tabular}{|c|c|c|c|c|}
\hline
 &GOP size (Kbytes) & number of GOPs\\
\hline
  Mean   &655.08 & 1262.79   \\
  \hline
   Standard Deviation & 201.44  & 271.46    \\
   \hline
   Standard error mean & 0.57 & 26.74 \\
   \hline
   Upper 95\% Mean & 656.20 & 1315.85\\
   \hline
   Lower 95\% Mean & 653.96 & 1209.74\\
   \hline
   Total no. of GOP & 124593 & 130068 \\
   \hline
   Min. GOP size  &1.91 & 580 \\
   \hline
   Max. GOP size & 2192.65 & 2018\\
   \hline
     
   \end{tabular}
   \caption{ statistics of GOP size  and number of GOPs in a video repository}
   \label{statistics}
   \end{table}

\subsection{View Pattern for Synthesized Videos} \label{view-model}
According to~\cite{sharma}, we model the access pattern to repository by applying Weibull distribution to generate long tail access pattern. The frequently accessed videos (FAVs) that have high access rates,  are located at the beginning of the Weibull curve  and those have less views are in the long tail.  We determine the percentage of FAVs number in the repository based on the shape and scale coefficients $\alpha$  and $\beta$ respectively of the Weibull distribution. As alpha decreases, the number of FAVs increases in the repository and vice versa. The interval range of $\alpha$ used in the experiment is $[0.4:2.4]$ while $\beta=1$


\subsection{Storage and Computation Prices}
The prices of storage and computation used in the experiments are modeled based on Amazon Web Service (AWS) prices as shown in Table~\ref{tab:price}
\begin{table}[ht]
\centering 
\begin{tabular}{c c c c c} 
\hline\hline 
 Transcoding & storage  & CDN \\ [0.5ex] 
\hline 
\texttt{t2-small} & \texttt{s3} & \texttt{CloudFront}\\
\hline
\$0.026  /hour & \$0.03   GB/month & \$0.085  GB/month \\ 
\hline 
\end{tabular}

\caption{AWS computation, storage, and Content Delivery Network (CDN) services and their prices in US dollar \$}
\label{tab:price}
\end{table}

\subsection{Baseline Methods for Comparison}
To evaluate our proposed algorithms, we compare them against two baseline methods that are \emph{Fully pre-transcoding} (that stores the whole video streams) and \emph{Fully re-transcoding} (that re-transcodes all video streams upon request).  These two methods ignore the metadata of processing and storage costs of the video streams.

\section { Experimental Results}
\subsection{Impact of  FAVs percentage in repository}
We create several synthesized repositories of video streams to illustrate the performance of the proposed algorithms. Each synthesized repository is made of total 50,000 videos with a percentage of FAVs. The shape parameter in of Weibull distribution ($\alpha$) controls the percentage of FAVs in the repositories. The percentage of FAVs  based on $\alpha $ and $\beta$ are shown in Table~\ref{FAVs}.  The average view of videos in the repositories is 1.99. We conduct each experiment  10 times and report the mean and 95\% confidence interval.
\begin{table}[ht]
\centering 
\begin{tabular}{|c| c c|} 
\hline
\multirow{6}{*} {$\beta=1$}&$\alpha$ &     FAVs  \\ [0.5ex]
\hline
& 0.4 &30\% \\ 
& 0.6   & 25\% \\ 
 & 1 & 20\% \\
 & 1.4 & 15\% \\
 & 1.8 & 10\% \\
 & 2.4 & 5\% \\
\hline

\end{tabular}
\label{price} 
\caption{Percentage of FAVs in the repository varies by changing $\alpha$ in the Weibull distribution}
\label{FAVs}
\end{table}

\begin{figure*}
\resizebox{2.21\columnwidth}{!}{%
 \begin{subfigure}[b]{0.5\textwidth}
\begin{tikzpicture} 
    \begin{axis}[
         width  = 0.9*\textwidth,
        height = 6cm,
        major x tick style = transparent,
        ybar,
        bar width=6pt,
       legend style={font=\fontsize{7}{5}\selectfont},
        ylabel = {total cost (\$)},
        xlabel={percentage of FAVs in the repository},
        symbolic x coords={5\%, 10\%, 15\%, 20\%,25\%,30\%  },
        xtick = data,
        scaled y ticks=real:1000,
    	ytick scale label code/.code={$\times10^3$},
        legend pos= north west,
           ]
       \addplot[area legend,ybar,fill=blue,error bars/.cd, y dir=both, y explicit] 
            coordinates {(5\%, 1596)+= (0,60) -= (0,23)(10\%, 1596)+= (0,60) -= (0,23)(15\%, 1596)+= (0,60) -= (0,23)(20\%, 1596)+= (0,60) -= (0,23)(25\%, 1596)+= (0,60) -= (0,23)(30\%, 1596)+= (0,60) -= (0,23)};

  \addplot[area legend,ybar,fill=red,error bars/.cd, y dir=both, y explicit] 
               coordinates {(5\%, 839)+= (0,29) -= (0,11)(10\%, 842)+= (0,29) -= (0,11)(15\%, 863)+= (0,30) -= (0,11)(20\%, 947)+= (0,33) -= (0,12)(25\%, 1424)+= (0,50) -= (0,19)(30\%, 3137)+= (0,111) -= (0,42)};

      \addplot[area legend,ybar,fill=green,error bars/.cd, y dir=both, y explicit] 
              coordinates {(5\%, 694)+= (0,24) -= (0,9)(10\%, 662)+= (0,23) -= (0,9)(15\%, 643)+= (0,23) -= (0,9)(20\%, 613)+= (0,22) -= (0,8)(25\%, 566)+= (0,21) -= (0,8)(30\%, 533)+= (0,20) -= (0,7)};

        \legend{Pre-transcoding Method, Re-transcoding Method, Partially Pre-transcoding Method }
         
    \end{axis}
   
\end{tikzpicture}

\caption{}
\label {favs}
\end{subfigure}%

 \begin{subfigure}[b]{0.5\textwidth}

\begin{tikzpicture}
    \begin{axis}[
         width  = 0.9*\textwidth,
        height = 6cm,
        major x tick style = transparent,
        ybar,
        bar width=6pt,
       legend style={font=\fontsize{7}{5}\selectfont},
        ylabel = {total cost including CDN (\$)},
          xlabel={percentage of FAVs in the repository},
        symbolic x coords={5\%,10\%,  15\%,20\%, 25\%,30\% },
        xtick = data,
         scaled y ticks=real:1000,
    	ytick scale label code/.code={$\times10^3$},
        legend pos= north west,
           ]
       \addplot[area legend,ybar,fill=blue,error bars/.cd, y dir=both, y explicit] 
            coordinates {(5\%, 6117)+= (0,229) -= (0,86)(10\%,  6117)+= (0,229) -= (0,86)(15\%,  6117)+= (0,229) -= (0,86)(20\%,  6117)+= (0,229) -= (0,86)(25\%,  6117)+= (0,229) -= (0,86)(30\%,  6117)+= (0,229) -= (0,86)};

  \addplot[area legend,ybar,fill=red, error bars/.cd, y dir=both, y explicit] 
             coordinates {(5\%, 839)+= (0,29) -= (0,11)(10\%, 842)+= (0,29) -= (0,11)(15\%, 863)+= (0,30) -= (0,11)(20\%, 947)+= (0,33) -= (0,12)(25\%, 1424)+= (0,50) -= (0,19)(30\%, 3147)+= (0,111) -= (0,42)};

      \addplot[ area legend,ybar,fill=green, error bars/.cd, y dir=both, y explicit] 
            coordinates {(5\%, 838)+= (0,29) -= (0,11)(10\%, 840)+= (0,29) -= (0,11)(15\%, 861)+= (0,30) -= (0,11)(20\%, 943)+= (0,33) -= (0,12)(25\%, 1234)+= (0,44) -= (0,17)(30\%, 1511)+= (0,56) -= (0,21)};
       
        \legend{Pre-transcoding Method, Re-transcoding Method, Partially Pre-transcoding Method }
         
    \end{axis}
   
\end{tikzpicture}

\caption{}
\label {cdn}
\end{subfigure}%

\begin{subfigure}[b]{0.5\textwidth}

\begin{tikzpicture}
    \begin{axis}[
         width  = 0.9*\textwidth,
        height = 6cm,
        major x tick style = transparent,
        ybar,
        bar width=6pt,
       legend style={font=\fontsize{7}{5}\selectfont},
        ylabel = {total cost (\$)},
        xlabel={ average number of views per video },
        symbolic x coords={1,2,3,4,5 },
        xtick = data,
         scaled y ticks=real:1000,
   		 ytick scale label code/.code={$\times10^3$},
        legend pos= north west,
           ]
        \addplot[area legend,ybar,fill=blue,error bars/.cd, y dir=both, y explicit] 
            coordinates {(1, 1596)+= (0,60) -= (0,23)(2, 1596)+= (0,60) -= (0,23)(3, 1596)+= (0,60) -= (0,23)(4, 1596)+= (0,60) -= (0,23)(5, 1596)+= (0,60) -= (0,23)};

  \addplot[area legend,ybar,fill=red,error bars/.cd, y dir=both, y explicit] 
               coordinates {(1, 947)+= (0,33) -= (0,12)(2, 1112)+= (0,37) -= (0,14)(3, 1451)+= (0,47) -= (0,18)(4, 1834)+= (0,60) -= (0,23)(5, 2234)+= (0,73) -= (0,27)};

      \addplot[area legend,ybar,fill=green,error bars/.cd, y dir=both, y explicit] 
              coordinates {(1, 769)+= (0,28) -= (0,10)(2, 543)+= (0,19) -= (0,7)(3,470)+= (0,16) -= (0,6)(4, 436)+= (0,15) -= (0,6)(5, 418)+= (0,15) -= (0,6)};

        \legend{Pre-transcoding Method, Re-transcoding Method, Partially Pre-transcoding Method }
         
    \end{axis}
   
\end{tikzpicture}
\caption{ }
\label {views}

\end{subfigure}%

}
\caption{Incurred costs of fully pre-transcoding, fully re-transcoding, and partially pre-transcoding methods: (\ref{favs}) when the percentage of FAVs varies in the repository; (\ref{cdn}) when the CDN cost is added to the cloud storage cost;  And   (\ref{views}) when the average view for videos in the synthesized repository is increased.}
\end{figure*}
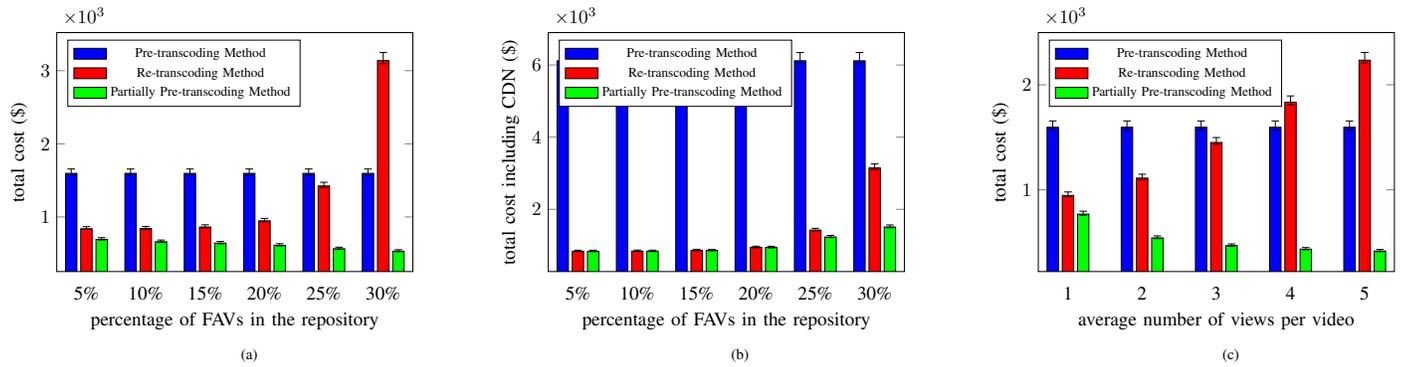

Fig. \ref{favs} shows the total cost of fully pre-transcoding, fully re-transcoding and partial pre-transcoding methods when the percentage of FAVs vary. When $\alpha$ is small, the percentage of FAVs increases and they, in turn, increase the cost of transcoding process.  As illustrated in the figure, the fully storage cost does not change and it stays constant  even when  changing the percentage of FAVs, because the fully storage cost is independent of the number of views of video streams.  

The experiment shows that our proposed method always reduces the incurred cost comparing to fully re-transcoding method by up to 70\% when 30\% of the repository is FAVs. Also, comparing to the fully pre-transcoding method, our proposed approach reduces the cost  up to 60\%. As the percentage of FAVs increases, our proposed method can reduce the total incurred cost significantly.

Similarly, in Fig. \ref{cdn},  we includes the cost of Content Delivery Network (CDN)  to the cloud storage cost. The proposed partial re-transcoding method reduces the cost up by 66\%. In fact, when we consider CDN, the total cloud storage cost is increased.  We note that as the cost ratio is increased, the number FAVs which are dependent on this ratio, decreases in the repository.

\subsection{Impact of average of number of views}
In this experiment, we increase the average number of views of FAVs without increasing percentage of FAVs, as shown in Fig. \ref{views}. In this experiment, the repository  includes 20\% of FAVs.  The average number of views of  FAVs is multiplied by constant 2, 3, 4, and 5 and the remaining videos  in the repository are divided by the same constants.

 The cost of fully re-transcoding increases, because FAVs have high  views which has direct impact on the re-transcoding cost. The proposed method outperforms the  fully pre-transcoding and fully re-transcoding methods. As the views of videos streams increases, the total cost of partial pre-transcoding method reduces.


\section{Related Work}


Gao \etal  \cite{gao} proposed a scheme that partially transcodes  video contents in the Cloud. Their approach aims to store the first segments of a video contents which are more frequently viewed, while transcode the remaining video contents online when requested, resulting storage and transcoding computation cost efficiency. Their system model is based on a partial transcoding of a video segments  and a dynamic storing system that is adapted to the video access rate change. They designed online algorithm to optimize the performance of their scheme. They stated that the viewers play 20\% of the video duration. they adopted the video access pattern by implementing the truncated exponential distribution to represent the video view rate. They demonstrated that their method reduces 30\% of operational cost by compared to storing the whole video.

Zhao \etal  \cite{zhao} proposed an approach to trade-off between computation and storage costs and to minimize the cost for multi version videos. They utilized the transcoding weight graph which is the transcoding relationships among versions of a video, along with the popularity of those different versions of the video. Based on the popularity and transcoding relationships among different video versions, their method decides which versions of a video should be stored in the repository or re-transcoded on demand. Their results shows reduction in the cost when compared to storing all versions.

 Jokhio \etal  \cite{jokhio} developed a strategy to strike a trade off between the computation and storage costs of a video. They estimated the computation cost, the storage cost, and the video popularity information of individual transcoded videos and then utilized  this information to make decisions on how long a video should be stored or how frequently it should be re-transcoded from a given source video. They compared their proposal to semisynthetic and realistic load patterns. Their results indicated that their strategy  is more cost-efficient than the two intuitive strategies.
\section{Conclusion}
In this paper, we propose a method to reduce the incurred cost of using cloud services through pre-transcoding, re-transcoding, or partially pre-transcoding of videos in the videos repository. We also analyze the impact of our proposed method when the rate of access to video streams in a repository varies. Experiment results show that as the percentage of FAVs in the repository  increases, the proposed method reduces the total incurred cost significantly. In particular, by increasing the number of views of FAVs, our method reduces the total incurred cost by up to 70\%.  Our future work will focus on  the prediction of the number of views of a video stream by applying machine learning techniques. The method  will operate based on views statistics over the past and thus we can develop offline algorithms to estimate  the cost of video stream transcoding in the cloud. This method will help VSPs to optimize the cost of using cloud resources.

\end{document}